# Geotree of Geodetector: An Anatomy of Knowledge Diffusion of a Novel Statistic

Yuting Liang[abc*], Jinfeng Wang[ab]

[a] *State Key Laboratory of Resources and Environmental Information System, Institute of Geographic Sciences and Natural Resources Research, Chinese Academy of Sciences, Beijing, China;* [b] *University of Chinese Academy of Sciences, Beijing, China; c Department of Human Geography and Planning, Utrecht University, Utrecht, the Netherlands*

*Corresponding author. Email: liangyt.19s@igsnrr.ac.cn.

**Abstract:** The growing number of citations to original publications highlighted their utility across academia, but the dissemination of knowledge from tacit conceptualization to scientific publications and its global applications remains understudied, and the prediction of knowledge trends in a disciplinary context is rare. Addressing the gaps, this paper constructed a tree-like hierarchical model (Geotree) to dissect the knowledge evolution paths of the Geodetector theory (a case) using the Web of Science citation database. Our results revealed that the knowledge evolution of 932 citations to Geodetector was partitioned into periods: a budding period of initial theoretical exploration, a growing period for emerging topics in application, and a mature period marked by significant citation growth. Our test $R^2$ of the predicting model over the next decade, considering the tree-like hierarchy across research directions and disciplines, was 100% higher than that of the other two (from 0.29 to 0.58). The knowledge spreading, from China to North America in 2011, Europe in 2012, Oceania in 2017, South America in 2018, and Africa in 2019, was more associated with a country's production of scientific publications (q-statistic = 0.307***) than its income level. The Geotree modeling of two other cases from space science and physics confirmed the reliability of the source publication-based approach in tracking knowledge diffusion. Our established research framework enriched the current methodology of information science and provided valuable references for policymakers and scholars to enhance their decision-making processes.

## 1. Introduction

Whether it is rumor propagation, social media posting, literature reading, or international conference dialogue, the flow of information is ubiquitous in both societal and academic realms (Yan, Zhou, Ren, Zhang, & Du, 2023). Knowledge dissemination serves not only as a carrier of information cascading, but also as an important catalyst for collaborative interaction among communities, organizations, and individuals (Lotrecchiano et al.). From the disseminator to the recipient, knowledge diffusion in previous studies was usually understood in terms of three dimensions (semantic content, time series, and geographical locations) that explained distributional heterogeneity and evolutionary trends of academic engagements. These dimensions and their dynamics were fundamentally generated by the learning mechanism of human domain knowledge. For example, as a core achievement of human domain knowledge, theories were continuously validated over time as they diffused among participants (F. Wang & X. Y. Wang, 2020). Naturally, cascading dissemination accompanies the transfer of knowledge and information from one layer to another or more layers, such as disciplines and regions. In this process, an adaptation, modification, and reinterpretation of theories often existed, while promoting knowledge sharing and win-win cooperation between connected domains (Semitiel-Garcia & Noguera-Mendez, 2012; Wuchty, Jones, & Uzzi, 2007). An intriguing question from the existing literature has been how domain-specific knowledge has cascaded to related and unrelated domains (Boschma, 2005; Wu, Dong, Wu, & Liu, 2023). However, how to represent, model, and predict the path of knowledge evolution, from data to methods to explanations, has remained an area of practice.

Understanding the modes and key factors of science communication is not only a theoretical pursuit but also a practical necessity to optimize the efficiency of information dissemination to break down disciplinary barriers. Existing studies revealed the process by which theories, concepts, or methods migrate over time from one domain to another (Parameswaran, Kishore, Yang, & Liu, 2023; Sun & Latora, 2020), offering opportunities for innovation while also presenting potential challenges associated with interdisciplinary integration. On one hand, the minor intersection of multidisciplinary theories, methods, and ideas can lead to significant innovation of science and technologies, creating a sustainable ecosystem for knowledge integration and evolution (Lee & Miozzo, 2019; Peng, Ke, Budak, Romero, & Ahn, 2021; Alexander M. Petersen, Dinesh Majeti, Kyeongan Kwon, Mohammed E. Ahmed, & Ioannis

Pavlidis, 2018; Sarin et al., 2020). Previous studies have demonstrated that the cascading dissemination of knowledge can benefit from high-impact journals, large author teams, comprehensive references, and detailed abstracts (Didegah & Thelwall, 2013; Mirnezami, Beaudry, & Lariviere, 2016; Nayak, 2022). On the other hand, certain niche knowledge of tools, software, and methods developed in lesser-known fields has struggled to gain acceptance across various domains (Bu, Lu, Wu, Chen, & Huang, 2021; Kaiser, Ito, & Hall, 2004; Kiss, Broom, Craze, & Rafols, 2010). One possible explanation is that without social networking connections and technological relatedness, scholars find it challenging to explore alternatives beyond their disciplines and apply them reasonably to their research (Balland & Boschma, 2022; Catalán, Navarrete, & Figueroa, 2022). Another possible reason is that scholars' disciplinary conservatism or disciplinary introversion, driven by familiarity and confidence in their own field's methods, limits cross-disciplinary cooperation and innovation. Publication pressures and disciplinary practices further reinforce their preference for familiar techniques (Alexander M. Petersen et al., 2018; Sinatra, Wang, Deville, Song, & Barabási, 2016; Wagner et al., 2021). While not entirely negative, disciplinary conservatism can impede multidisciplinary exchange and innovation in regions (Autant, Mairesse, & Massard, 2007; Bretschger, 1999).

Previous scholars have made unremitting efforts in this regard. Considering the early exploration of theory diffusion, Kaiser et al. (2004) published an article titled "Spreading the Tools of Theory: Feynman Diagrams in the USA, Japan, and the Soviet Union". Feynman diagrams were initially developed as a tool for intricate computations, and their comprehension and absorption took some time of familiarization and apprenticeship. A subsequent study successfully applied epidemiological models to the spread of Feynman diagrams to demonstrate quantitative similarities between the spread of ideas and infections (Bettencourt, Cintron-Arias, Kaiser, & Castillo-Chavez, 2006). However, the current body of such research is inadequate, and the multidimensional process of knowledge diffusion from specific domains to broader spheres, as well as its future trends remains insufficiently elucidated. Subsequently, scholars are gradually recognizing the structural diversity and phased nature of knowledge processes, such as network structure (C. Liu, Shan, & Yu, 2011; Semitiel-Garcia & Noguera-Mendez, 2012), diffusion paths (Lu & Liu, 2013; Yu & Sheng, 2021), citation relationships (Ba, Ma, Cai, & Li, 2023; M. Kim, Baek, & Song, 2018; Savin, Ott, & Konop, 2022), knowledge flow matrix (Ko, Yoon, & Seo, 2014), and the core-peripheral structure (Z. F. Chen & Guan, 2016). Concurrently, some studies emphasize the validity of the evolutionary perspective in analyzing knowledge diffusion, such as the cases of population dynamics (Huang, 2013) and R&D networks (Morescalchi, Pammolli, Penner, Petersen, & Riccaboni, 2015). Thereafter, scholars have revealed diffusion stages of knowledge through empirical studies. Notably, Wang et al. defined four modes of theoretical construction: elaboration, proliferation, competition, and integration (F. Wang & X. Y. Wang, 2020); Sun et al. identified the diffusion behavior of specific domain knowledge (modern physics) over time, including its absorption by other fields, mutual influences, and reciprocal nourishment (Sun & Latora, 2020); Parameswaran et al. categorized the diffusion of IT innovation knowledge into four stages: emergence, structurization, evolution, and chasm (Parameswaran et al., 2023). Despite being incomplete, there are increasing empirical studies and theoretical underpinnings for determining the mechanisms and upcoming dynamics of knowledge evolution.

To sum up, the multi-disciplinarity and cross-disciplinarity of knowledge are critical to scientific advances and innovation. The current literature has contributed to the dimensions, process, structure, and stages of knowledge diffusion using publications and patents. Despite significant advancements and challenges in understanding knowledge diffusion, the literature still grapples with the intricate question of how knowledge cascades into related and unrelated domains of science. To answer this question, our study aims to track the knowledge diffusion process from the source publication to cascading citations across disciplines worldwide. By doing so, the source publications will characterize the original knowledge, their cascading citations convey the knowledge diffusion footprints, and the semantic topics depict the scientific interests of scholars.

In this study, we used a case from an evolutionary perspective to achieve our research goals. The dynamic process in which knowledge in a specific field cascades from scratch to many fields around the world was illustrated by the theory application and development history of a geographical tool, Geodetector, which was officially published in 2010 (Jin-Feng Wang et al., 2010). The reason for choosing Geodetector is that this tool from the field of geographic information science has rapidly formed many cross-disciplinary research cases around the world, providing historical data for the issues we intend to explore (Li et al., 2020; Yin, Wang, Ren, Li, & Guo, 2019). Therefore, 932 citations of Geodetector were collected from the Web of Science (WoS) database as our research data. The application of Geotree software in our research provided a novel and comprehensive solution to the appealed problem. Based on the construction of the evolutionary tree model, we analyzed the different diffusion stages of the Geodetector theory based on their citation growth rates, future trends, and influencing factors. Moreover, to validate the diversity of other knowledge diffusion patterns within the framework of a tree-like evolution structure, we briefly discussed the modeling results of knowledge evolutionary trees for the diffusion of two other theories: one from spatial science (Anselin, 1995) and the other from physics (Einstein, 1905). Tracing such diffusion processes across multiple dimensions helps understand the dynamic flow patterns and knowledge trends in different environments. This understanding provides references for organizations, social networks, and online platforms to develop more precise information dissemination strategies, to enable information to penetrate target groups more quickly and effectively.

## 2. Materials and methods

### *2.1 Dataset and pre-processing*

Once scholars had cited previous studies, the potential paths for science communication and knowledge dissemination became traceable and calculable. To provide a more comprehensive description of the dissemination process, we collected relevant citation data from all collections using a web crawler (Python) rather than exporting the core collection of the Web of Science database (WoS). On January 11, 2021, 1331 citations, limited to articles published between 2010 and January 11, 2021, were downloaded from WoS (**Table 1**). Geodetector was proposed in the listed three source publications, laying its foundation in terms of principle, measurement, and practice. Their citations reflect the diffusion footprints of Geodetector knowledge in WoS. Among them, 567, 308, and 456 citations were retrieved from documents A (Jin-Feng Wang et al., 2010), B (J. F. Wang, Zhang, & Fu, 2016), and C (J. Wang & Xu, 2017), respectively. 932 documents were used as experimental data after the deduplication processing and exclusion of the documents of the year 2021. The dataset included seven fields, namely, article URL, title, publication year, abstract, author address, WoS classification, and research area. The latitude and longitude of each citation were extracted through toponym matching and geocoding processing of the field of “author address”.

**Table 1 Data sources: three source publications and their citations of Geodetector**

| Document | Published year | Title | Journal | Content | Citation counts in WoS |
|---|---|---|---|---|---|
| A | 2010 | Geographical detectors-based health risk assessment and its application in the neural tube defects study of the Heshun Region, China | International Journal of Geographical Information Science | Nonlinear attribution and interactions | 567 |
| B | 2016 | A measure of SSH | Ecological Indicators | A measure of SSH and PDF of SSH | 308 |
| C | 2017 | Geographic Detector: Principles and Prospects | Acta Geographica Sinica | principle, typical applications, and Q/A | 456 |

Notes: SSH refers to spatial stratified heterogeneity; PDF refers to the probability density function

***2.2 Thematic analysis: Latent Dirichlet Allocation***

To analyze the different paths of knowledge diffusion in citations, a classical topic classification model, Latent Dirichlet Allocation (LDA), was introduced to identify the disciplines and research directions of the citation documents (**Fig.1 A**) (Blei, Ng, & Jordan, 2003; Deerwester, Dumais, Furnas, Landauer, & Harshman, 1990; Hofmann, 1999; Ramage, Hall, Nallapati, & Manning, 2009). Suppose a corpus $D$ consists of $M$ documents, with document $d$ containing $N_d$ words ($d \in 1, \ldots, M$). The probability of a corpus was calculated as follows (Blei et al., 2003):

$$p(D|\alpha, \beta) = \prod_{d=1}^{M} \int p(\theta_d|\alpha) \left( \prod_{n=1}^{N_d} \sum_{z_{dn}} p(z_{dn}|\theta_d) p(w_{dn}|z_{dn}, \beta) \right) d\theta_d \qquad (1)$$

Here, the Stanford Topic Modeling Toolbox (TMT) of the Stanford Natural Language Processing Research Group, first proposed in September 2009 (Ramage et al., 2009) (https://downloads.cs.stanford.edu/nlp/software/tmt/tmt-0.4/), was used to carry out Latent Dirichlet Allocation (LDA) topic analyses.

To reflect the evolutionary path of multidimensional knowledge in a primary class (discipline) and its secondary class (research direction) (Liang & Xu, 2023; J. F. Wang et al., 2012; Yuting, Yunfeng, & Yueqi, 2019), we need to process the generated LDA topics. In this paper, two experiments on topic classification were designed. The criteria for the number of topics were to increase or decrease the number of topics, while the generalized topics were no longer increased or decreased. We first divided the citations into ten LDA topics and then into 30

LDA topics. According to the statistics of calculated probabilities of generated topics (e.g. the 30 topics in **Fig.1 B**), 90% of documents were significantly more likely to be attributed to the first topic than the second, so the probabilistic first topics were consistently identified as the document's topics. The ten generated topics were assigned into five first-level categories in terms of the corresponding key terms. The names of these five first-level categories, Geosciences (GS), Agricultural Sciences (AS), Health Sciences (HS), Mathematics and Statistics (M&S), and Atmospheric Science & Meteorology (AS&M) (**Fig.1 C**), were referencing to the discipline explanations of the National Science Foundation (NSF) Codes for Classifications for Research (https://osp.unm.edu/pi-resources/nsf-research-classifications.html). Then, among the 30 topics, topic 2 was automatically identified as the "other category." The five documents in the topic were classified under the first-level categories closest to their abstract contents, so the final quantity of topics for second-level categories (research directions) was corrected to 29 (**Fig.1 D**). Finally, each document belonged to one unique discipline type and one unique research direction type.

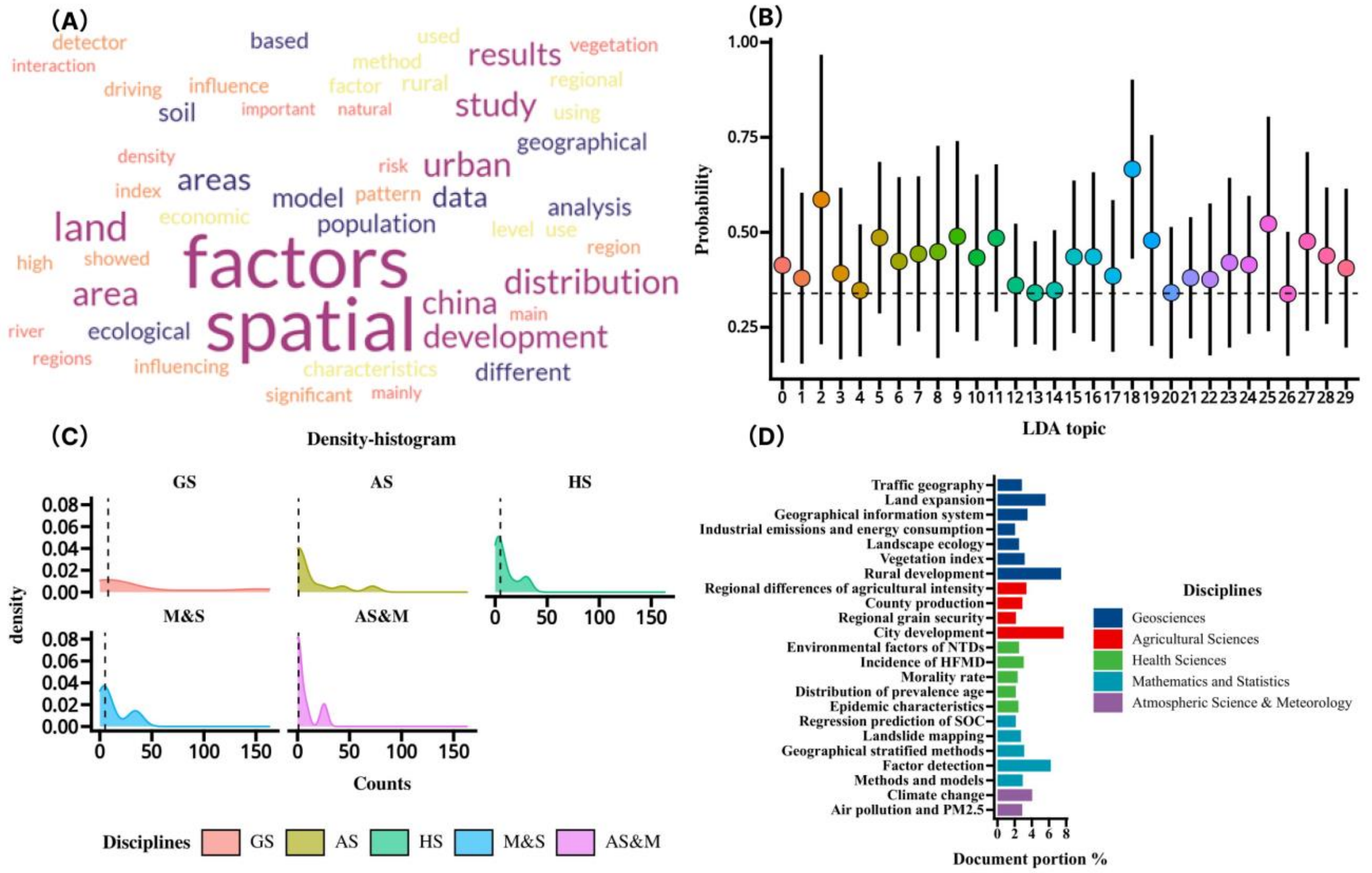

**Fig.1. The results of thematic analysis of Geodetector citations. (A):** Word cloud picture for all abstracts of Geodetector citations; **(B):** Probability distribution of all LDA topics; **(C):** Density

histogram of yearly citation count with facets in five disciplines; **(D):** Document distribution for the final research.

### *2.3 Geotree: an evolution tree model*

How can textual data be projected as knowledge diffusion paths? We introduced the Evolutionary Tree Model, a multi-dimensional coordinate system centered on natural tree evolution, to model knowledge diffusion paths, offering a simple visualization for complex phenomena and diversity (J. F. Wang et al., 2012). Understanding the research object's growth pattern via biological evolution allows inferring its past or future state from a single observation, exchanging time states for attribute space (Jing, Wang, Xu, & Yang, 2022). Previous empirical cases from different disciplinary backgrounds have provided a scientific foundation and reference experience for constructing theoretical diffusion models of knowledge evolution, such as non-communicable diseases (Y. Wang & J. Wang, 2020), urban land expansion (Jing et al., 2022; J. F. Wang et al., 2012), and literature reviews (Duan et al., 2020; Liang & Xu, 2023). This model necessitates unique categorical and stage variables, enabling the construction of the knowledge evolution tree structure while presenting challenges in integrating citation data.

To construct the tree-like structure, we modeled the taxonomic evolutionary process of citations using Geotree software (http://www.sssampling.cn/geotree/GeogTree), a comprehensive new statistical analysis software whose theoretical basis is the evolution tree model. The final evolution tree structure was established and visualized using Draw.io (https://app.diagrams.net/). The modeling of the knowledge evolution tree model was performed according to the following steps.

#### *2.3.1 The construction of discipline-research direction trees*

First, to describe the cascading distribution of citation content in different disciplines and research directions and its characteristics over time, we constructed discipline-research direction trees by years. Based on the thematic analysis results in section 2.2, the five discipline types and 29 research directions were successively inputted as the first-level branches and second-level twigs of the Geotree, and a leaf represented a citation. In total, modeling results for ten years, 2010~2015, 2016, 2017, 2018, 2019, and 2020, are included.

*2.3.2 The construction of the knowledge evolution tree*

Then, to explain the evolution process of knowledge diffusion and its future trends, we constructed the Goetree of Geodetector. The five discipline types were set as the five first-level branches. The three development stages of budding, growth, and maturity were incorporated as the second-level twigs of the Geotree structure, and a leaf represented citation counts per research direction.

*2.3.3 The construction of the factor evolution tree*

Finally, an evolutionary tree was constructed to analyze the impact of economy type and development stage on knowledge diffusion in different regions. Economy types and development stages corresponded to the branches and twigs of the evolution tree, respectively, and a leaf represented all citations of an economy. Economy types were based on the World Bank Atlas Method, World Bank Country, and Lending Groups for the 2021 fiscal year. The development stages were based on the Science and Engineering Indicators of 2020, calculated as the ratio of the fractional count of science and engineering (S&E) articles in all S&E fields by country or economy in 2020 to the global total. Therefore, the economies were divided into four income types: high, upper-middle, lower-middle, and low. Three development stages were less than 1%, 1%–3%, and more than 3%.

## 3. Results

To analyze the evolutionary pathways of knowledge diffusion, our study started from citation growth to topic branches to branch evolution, future predictions, and their influencing factors. Therefore, we compared the different citation growth trends for three source publications of Geodetector. Against those dynamic trends, we constructed the evolution tree structure to illustrate the historical evolutionary pathways and their development trends in the next ten years. Connecting the coordinates of these citations, we identified areas of under-adoption and studied the possible influencing factors. Finally, the modelling to two other cases in space science and physics were performed, which proved the practicability of our approach from citation documents to knowledge evolution trees.

### *3.1 Tree-like knowledge evolution processes*

*3.1.1 Modeling of knowledge evolution structure*

The research case used to analyze knowledge diffusion is Geodetector, proposed in 2010. From the conceptualized year (2010) to 2020, the traced knowledge diffusion has differentiated into primary pathways in five disciplines and secondary pathways in 29 research directions. For the topic branches, the diffusion paths were illustrated by six discipline-research direction trees with primary branches as disciplines and secondary twigs as research directions for the 2010–2020 time period. For the branch evolution trends, we defined three stages by calculating the diffusion rates in terms of new citation counts per year per research direction ($Diffusion\ rate_{i,t} = \sum_{n=1}^{n=k} New\ citation_{i,t}$, k is the number of new citations, i represents research direction and t represents year). Among the diffusion rates of each research direction over eleven years, the minimum was 0, median was 3, and maximum was 45. A diffusion rate of 0 indicated that the research direction has not yet been created; a diffusion rate equal to 1 indicated the creation of the research direction; and a diffusion rate greater than 1 indicated that the research direction was in the development stage. Similarly, relevant previous studies on technology development and stages showed that knowledge diffusion in a field would indicate early and late stages with less diffusion due to the limited number of early knowledge disseminators, adoption barriers, and market saturation (Menanteau & Lefebvre, 2000; Methé, 1992). Therefore, we classified the years with diffusion rates between [0,3] as the first stage, which indicated the early budding stage of knowledge creation; the years with research directions with diffusion rates greater than 3 and accompanied by the emergence of new research directions as the second stage, which indicated the growth stage of diffusion; and the years with diffusion rates greater than 3 and no emergence of new research directions as the third stage, which indicated the mature stage of diffusion.

Among the six tree-like structures, **Fig.2 A** illustrates the budding stage of the Geotree including years from 2010 to 2015, branched by the disciplines and research directions, in which the first-level branches were arranged by the year sequence from the old to new. As shown, the largest branch was GS, indicating that there were many applications in GS but few in AS&M and AS. Scholars in each branch developed different twig pathways of scientific interests. (1) From the twigs of GS, Geodetector has been applied to nine research directions. Among them, the most cited were geographic information system (GIS) and rural development, and the least cited were traffic geography and landscape ecology. (2) From the twigs of AS, Geodetector has been applied to three research directions. (3) From the twigs of the AS&M, only one citation was related to climate change during the early five years. (4) From the twigs of HS, scholars were concerned with six research directions, such as the environmental factors of neglected tropical diseases (NTDs), and the incidence of hand, foot, and mouth disease (HFMD). Among these, most citations were about the distribution of the prevalence age. (5) From the twigs of M&S, there were five research directions, such as soil organic carbon (SOC) regression prediction, and geographical stratified methods (GeoStratified). Among them, most citations were for GeoStratified research.

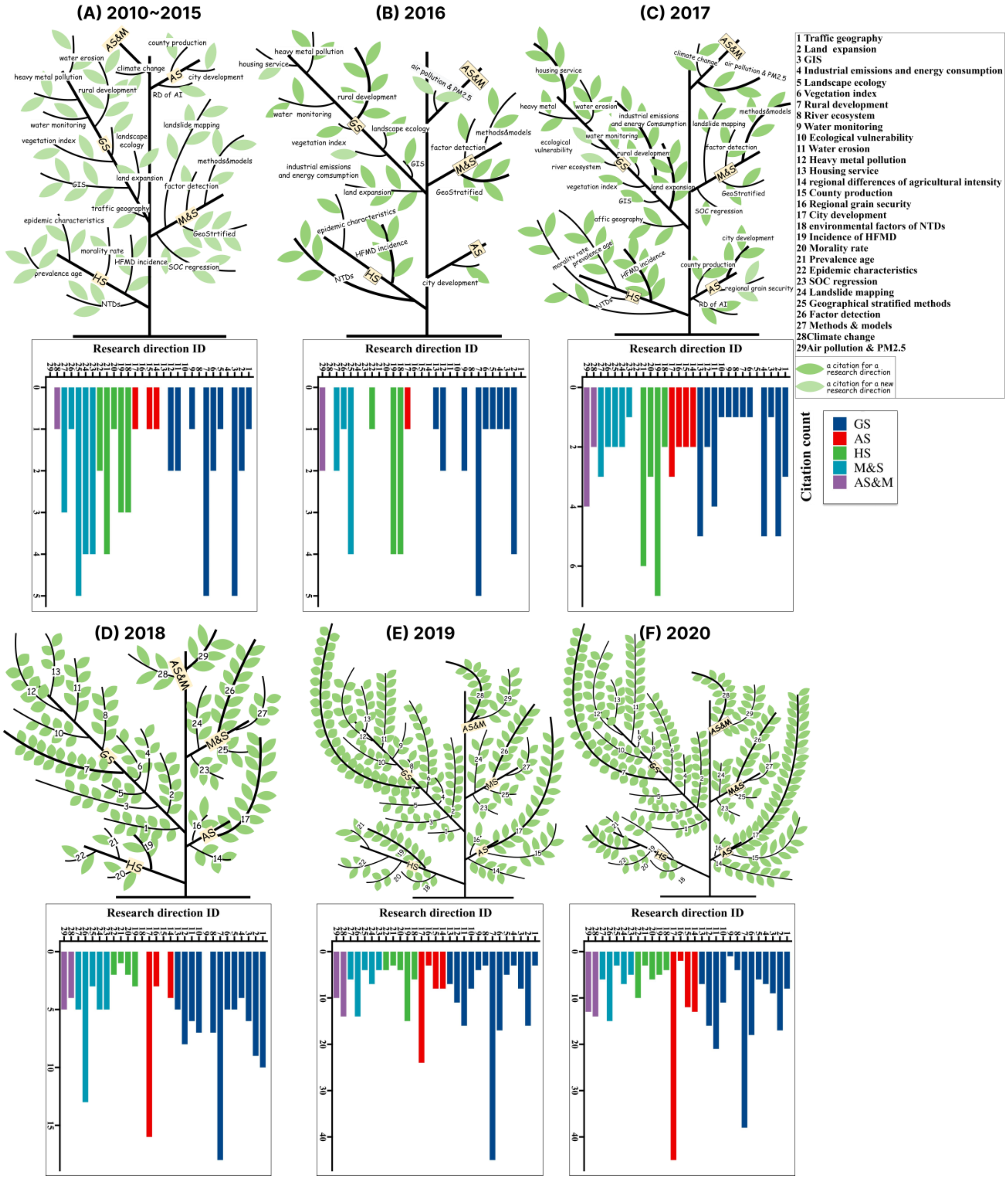

**Fig.2. The discipline-research direction trees from 2010 to 2020.** The modeling data are Geodetector citations indexed in Web of Science. The first-level branches of each tree represent disciplines. The second-level twigs represent research directions. Each leaf represents a citation of Geodetector.

**Fig.2 B** is the first growing stage of the Geotree in 2016. Following the software release of Geodetector in 2015 (www.geodetector.cn) and the second representative paper published in 2016, the citation counts increased significantly, and dominant disciplines appeared. **Fig.2 C** is the second growing stage of Geotree in 2017. After the third source paper was published in 2017, the citation counts of the three source articles reached 73, an increase of 36 compared to 2016. From 2016 to 2017, there were three new emerging research directions for GS (industrial emissions and energy consumption, housing service, and food security) and one for AS&M (PM2.5). **Fig.2 D, E,** and **F** represent the mature stage of Geotree from 2018 to 2020. Starting from 2018, there were no new emerging research directions in each discipline, and the types of research directions in 2019 and 2020 were the same. Although the first paper triggered the most significant overall citation growth, subsequent representative papers and software effectively promoted the staged increase in citations from the budding stage to mature stage (Furman & Teodoridis, 2020).

Combining Geotree's type variables and state variables can facilitate the study of the branched evolution of research objects. Based on the thematic categories and diffusion stages of citations, the Geotree of Geodetector model from 2010 to 2020 was constructed to depict the knowledge evolution pathways (**Fig.3**). According to the emerging sequence of different disciplines, the earliest scientific interests in the field of Geodetector was found among the scholars from HS (2010) and M&S (2010), followed by GS (2011), AS (2011) and AS&M (2015). In the budding stage, fewer than five citations existed in each research direction; in the growing stage, the growth of citations accelerated, and many new research directions emerged; in the mature stage, the number of new citations per year in each research direction except R18 exceeded five, and the citation count in research directions was also the highest. Although the citation distribution and data volume of five disciplines varied greatly, they all demonstrated a phased trend of knowledge development at the same points in time; the emergence, development, and prosperity processes for each research direction had different paths. The tree-like model intuitively integrated the complex multi-type and multi-state data to a hierarchical structure, allowing us to analyze the cascading paths of knowledge diffusion by discipline and research direction.

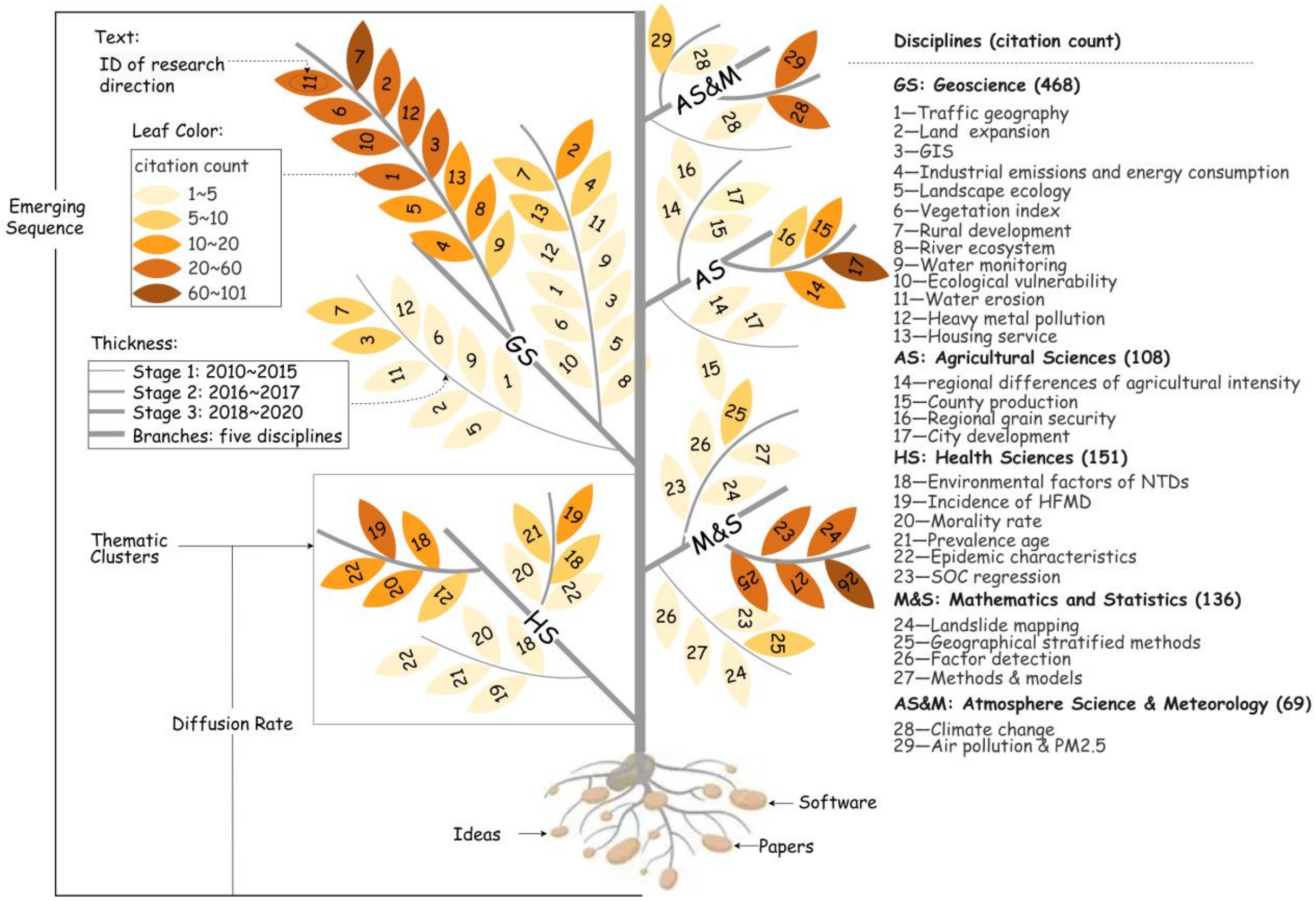

**Fig.3. The knowledge evolution tree from 2010 to 2020.** The modeling data are Geodetector citations indexed in Web of Science. The first-level branches represent five disciplines. The second-level twigs represent three diffusion stages. Each leaf represents a research direction.

### *3.1.2 Prediction of knowledge evolution paths*

Predicting citation counts is crucial as it not only gauges the prospective impact and relevance of research but also guides strategic decision-making in academia and policy. In our study, we compared the Linear Regression (LR) and Multi-Level Model (MLM), subjecting both to rigorous cross-validation to ensure the robustness and reliability of our findings. First, the data from 2010 to 2020 were randomly divided into ten subsets, nine of which were randomly selected as training subsets for model training, and the remaining test subset was used for accuracy verification. Then, the modelling was cross-validated 100 times for each fold (a total of 1000 iterations). Finally, we compared key metrics including R-squared ($R^2$), Root Mean Square Forecasting Error (RMSFE), and Mean Absolute Forecasting Error (MAFE) on both training and testing sets derived during the cross-validation (Savin and Winker, 2013).

Our results indicated that the LR2 (R equation: annual new citation count ~ year) generally outperformed the LR1 (R equation: annual new citation count ~ year) and MLM (R equation: annual new citation count ~ year + (year | discipline) (**Fig.4 A**). The difference between LR1 and LR2 was that the input of LR1 involved historical data of all research directions, while LR2 modeled the data of each research direction individually. The $R^2$ values of LR2 on the training and testing sets were 0.6 and 0.58, respectively, indicating good model performance, can reveal the variability of the target variable, while small RMSFE and MAFE values indicated that the prediction error of the model was relatively low. Although the MLM model improved the accuracy of the training set compared to LR1, it cannot generalize well to new data. This shows that different disciplines had widely different citation distribution patterns, causing the model to focus too much on some categories and perform poorly on others, thus not capturing the complex relationships between disciplinary features effectively. Even if other complex factors were not considered, the annual number of new citations of the Geodetector software in different research directions had a significant linear relationship with the time variable. All results passed the cross-validation. Although the $R^2$ of different research directions was different, the average values of the trained and tested sets were relatively close (**Fig.4 B**). Therefore, we utilized the LR2 to predict the number of new annual citations within 29 research directions for the upcoming ten years. Compared with the other two models, the test R2 of LR2 has improved 100% (from 0.29 to 0.58).

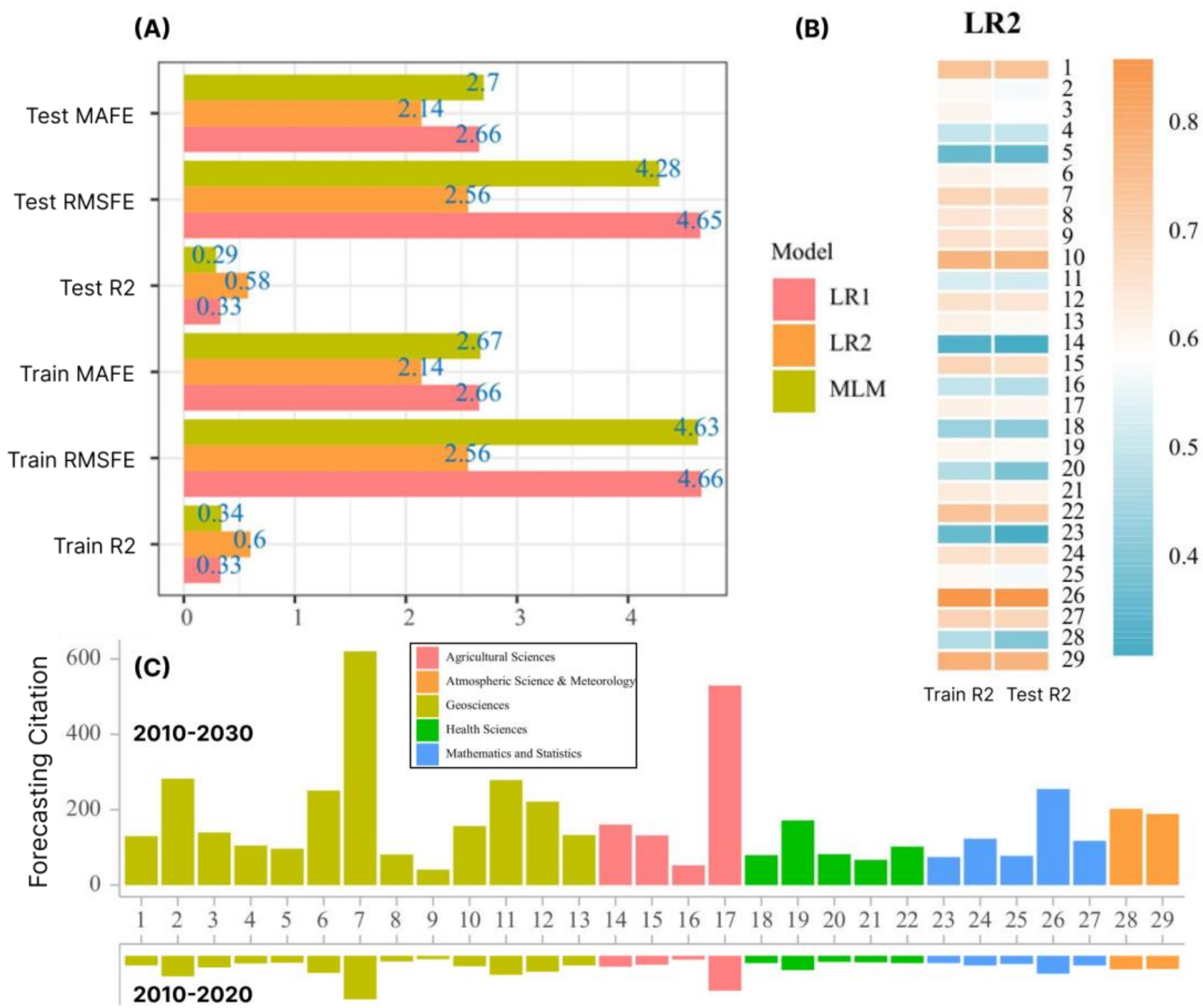

**Fig.4. The results of LR and MLM models for trained and tested data. (A):** the comparison of modeling performances between LR and MLM; **(B):** the trained $R^2$ and tested $R^2$ of LR prediction for each research direction; **(C)** the prediction of citations in each research direction in 2030.

By using the number of new citations, we focused on observing specific changes at each point in time, independent of the cumulative sum of previous ones. This approach helped to more accurately capture trends and cyclical changes without being distorted by historical cumulative values. The number of new additions per year was predicted, and the total number of citations in the target year was obtained after accumulation. As shown in **Fig.4 C**, the cumulative number of citations in 2030 is increasing in all 29 research directions. Referring to the trends observed so far, the application of Geodetector in rural development (R7) and city development (R17) will reach significant peaks. This phenomenon suggested that the advantages of certain fields of application expand as the number of studies increases. However, the research related to water monitoring (R9) will remain at a minimum level, suggesting a sustained increase in the total number of citations won't improve the shortcomings in specific fields. Notably, the research content also influenced the citation growth under both trends. For instance, the literature related to R7 and R17 focused on the practical inspirations obtained from applying Geodetector, while the scholars involved in R9 focused on improving related theories and models in their fields. The application in regional grain security (R16) will also be maintained at a very low level, indicating areas where the theory of Geodetector has been introduced only episodically and not become mainstream.

### *3.2 Uneven distribution of global knowledge diffusion*

#### *3.2.1 Mapping of geospatial and disciplinary spaces*

For the dimension of geographic locations, our results show that articles citing Geodetector have been published by scholars from 25 countries **(Fig.5 A)**. Globally, China, as the birthplace, produced the most citations (870), the United States produced 18 citations, Australia and Italy each produced five citations, four countries (South Africa, Canada, Germany, and Iran) each produced three citations, and the remaining 17 countries produced one or two citations. In China, 50 cities each produced 1–5 citations, seven cities each produced 6–10 citations, 12 cities each produced 11–20 citations, six cities each produced 21–50 citations, and Beijing produced 240 citations. According to the publication years of the first citation in different places, we found that knowledge of Geodetector first spread from north to south and east to west from the city (Beijing) to other cities in China. It then spread to several other continents at different times, including North America (2011), Europe (2012), Oceania (2017), South America (2018), and Africa (2019). Globally, it spread from east to west and then from north to south.

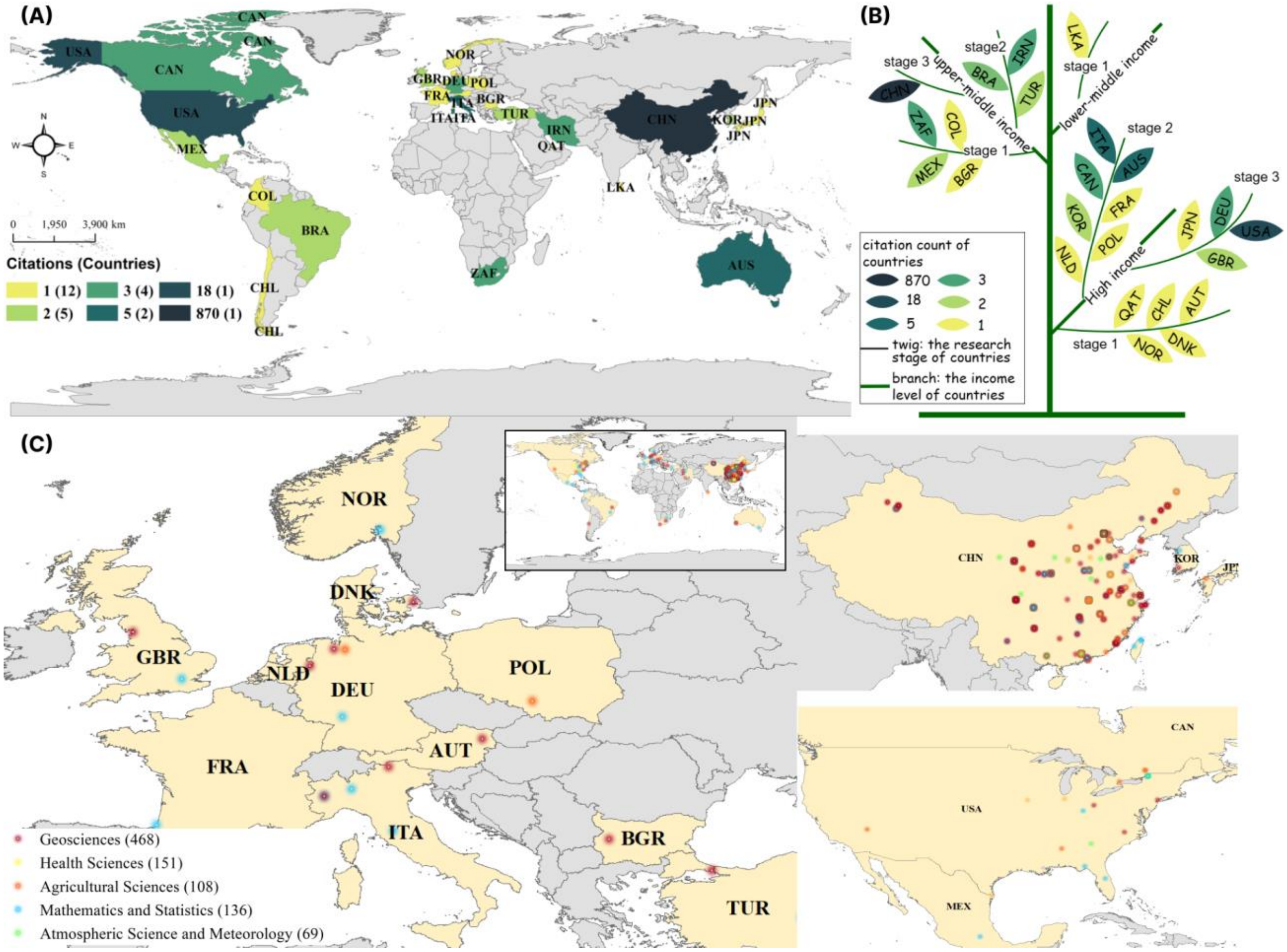

**Fig.5. The spatial patterns of the citing authors of Geodetector. (A):** The map of Geodetector citations in the world. The color fill of each country represents the number of citations published by authors in that country, and the corresponding number in parentheses represents the number of countries with that number of citations. **(B):** The factor evolution tree of Geodetector theory; **(C):** The distribution of Geodetector application disciplines. The color of the dots represents the subject type of the citations, and the corresponding number in parentheses represents the corresponding number of citations.

For the pattern of disciplines and adoption (**Fig.5 C**), we found that Geodetector has not yet been adopted by South America, Africa, and Australia in agricultural sciences, health sciences, and meteorology, while unadopted areas in Europe, include health sciences and meteorology. In most cases, this was caused by the disparity between the scientific research inputs and outputs of the different economies. According to the National Center for Science and Engineering Statistics (NCSES), approximately 85% of global S&E articles were reported in the eight largest scientific fields. For example, in terms of the disciplinary distribution of S&E article output in China and India in 2020, scholars of engineering published the most articles (15.41% and 16.15%, respectively), and social sciences published the least (1.28% and 1.71%, respectively). As for the United States, Japan, the United Kingdom, and some other European countries, the health sciences subject was always in a leading position (for example, 36.62% in the United States, 32.30% in Japan, and 33.32% in the United Kingdom), materials science was the least developed field, and social sciences were in the middle. In contrast to developed countries, developing countries had different models of scientific development, which has led to significant inequities in development.

Although Geodetector knowledge has been cascaded from a niche field into many fields for the analysis of spatial heterogeneity and influencing factors, including public health (Li et al., 2020; Liao et al., 2013), soil science (Yang et al., 2021), social science (B. Chen, Song, Kwan, Huang, & Xu, 2018; Zhan, Kwan, Zhang, Wang, & Yu, 2017), and statistics and environmental metrology (Broadbridge, Kolesnik, Leonenko, Olenko, & Omari, 2020; Fattorini, Marcheselli, Pisani, & Pratelli, 2020). According to the records of Web of Science (WoS) (retrieval date: July 10, 2023), this theory has been cited by papers from 88 WoS categories and 61 research directions affiliated with 1013 institutions from 54 countries worldwide, involving 3353 scholars and 1090 funding institutions. Despite such extensive international and interdisciplinary diffusion of the theory, some scholars in economic geography, human geography, and remote sensing science were unaware of this theory, let alone scholars outside the field of geography, highlighting the limitations in the theory diffusion process. Such limitations may arise from intertwined factors, such as disciplinary conservatism, geographical boundaries, and variations in the quality of scientific publications (Autant et al., 2007; Bretschger, 1999).

### *3.2.2 Tree evolution of influencing factors*

To quantify the influencing factors for each relevant economy, we constructed a knowledge factor evolution tree model consisting of economy types and scientific research stages. All countries cited in the Geodetector research were divided into three stages according to the percentage of their total number of S&E publications in 2020. As illustrated in **Fig.5 B**: (1) among high-income countries, several countries in the first stage had similar knowledge adoption rates for Geodetector; among the seven countries in the second stage, Italy and Australia were significantly higher than France; and among the four countries in the third stage, the United States was significantly higher than Japan. (2) Among the upper-middle-income countries, South Africa in the first stage was much higher than Colombia and Bulgaria; Iran, Brazil, and Turkey did not differ much in the second stage; and China in the third stage had an absolute advantage in terms of quantity. (3) Among the lower-middle-income countries, Sri Lanka was in the first stage and had a small number of publications. Based on the current research data and results, Italy has adopted more Geodetector knowledge than France, which had a higher economic level than it; South Africa has adopted more than Bulgaria, which had a higher economic level. However, Geodetector knowledge has not yet been adopted by low-income countries based on our present data.

Geodetector q statistic was used to attribute the dependent variable to the independent variable; that is, 100*q% of the variance of the dependent variable was explained by the independent variable. Our analysis of the Geodetector q statistic of Geodetector citations by national types and scientific research stages demonstrated that, when excluding China, q=0.146*** (p-value<0.01) and otherwise q=0.307*** (p-value<0.01), the scientific research stages had significant explanatory powers for the citations of Geodetector. Therefore, the knowledge diffusion rate in the cases of this study had a significant nonlinear correlation with S&E article output, while the economy types had a negligible impact on citation growth.

Thereafter, our attention shifted to this niche software from the GIS area of geography. We, in turn, set dependent variables as scientific research stages that reflect science and engineering (S&E) articles in all S&E fields by country or economy in 2020 to the global total, which mean they inherently contained the external and mixed context of the global academic environment beyond the local variance of Geodetector knowledge diffusion. We set Geodetector citations as the independent variable to represent niche knowledge diffusion. The results show that Geodetector's knowledge diffusion had significant impacts on the global academic community in all three stages (stage 3: q-statistic = 0.146***; stage 2: q-statistic = 0.151***; stage 1: q-statistic = 0.149***). Compared to the first and third stages, the second stage, which had the fastest dissemination of knowledge and most emerging cross-cutting research topics, had the highest global impact. Although the explanatory power of the individual stages was relatively small, the changes caused by the theoretical diffusion from a small field were all significant at 95% confidence intervals for the global context.

### *3.3 The generalization of knowledge evolution tree*

*3.3.1 The source publication-based approach*

To apply Geotree modeling in knowledge diffusion for any research topic, there are three main steps to follow (**Figure 6**). Firstly, we input the source publication of targeted knowledge in the search query of Web of Science database. Therefore, the citation data of this representative paper will be downloaded for free in 'plain text' format. Secondly, input the text data into thematic analysis modeling. Here we are using the TXT tool for LDA algorithm to generate research topics. Thirdly, the topic classified data will be prepared to input into Geotree software. We select the first-level categories (topics) as the branches, the second-level categories (years) as the twigs, and each citation will be as a leaf on the tree. All generated tree structures can be reproduced in other visualization tools (e.g. Draw.io and Adobe Photoshop).

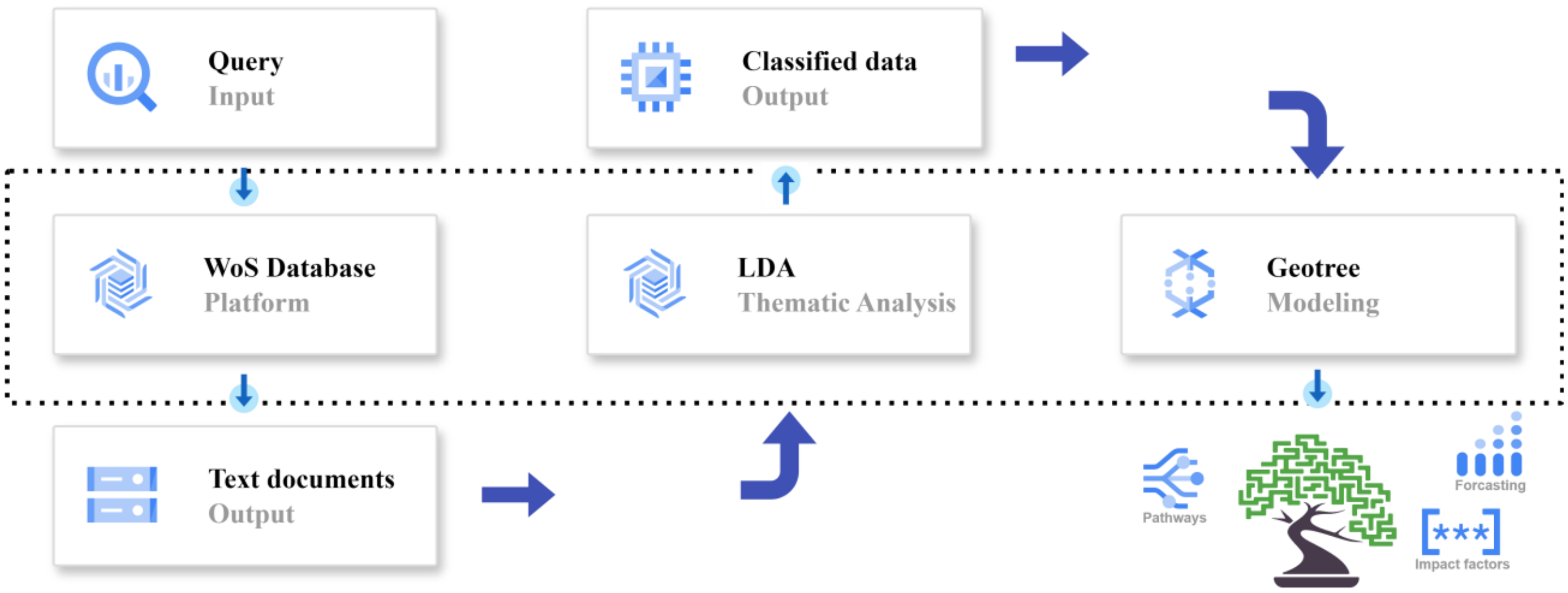

**Figure 6 Modeling knowledge diffusion with Geotree and citation data**

*3.3.2 The evolution tree of LISA*

Finally, to generalize the approach of tree-structured knowledge diffusion, the Geotree of local indicators of spatial association (LISA) referring to the Geotree of Geodetector was constructed (see **Fig.7 A**). On March 7, 2021, 5370 citations of the source publication were retrieved, which was titled "Local indicators of spatial association—LISA," which was first proposed in 1995 by Luc Anselin (Anselin, 1995). Finally, 4732 records for LISA (1996–2021) were downloaded from the WoS database. Among them, 4646 records for LISA were downloaded from abstracts. After excluding the citations in 2021, 4436 were used for the resulting tree of LISA (from 1996 to 2020). Considering the large number of citations for LISA, each leaf represented a certain number of citation collections rather than a single citing article. Among the citations of LISA, 743 are related to health sciences, 1697 to social sciences, 277 to atmospheric science and meteorology, 1067 to mathematics and statistics (mainly methods and models), and 652 are related to geosciences.

All the above citations are divided into five stages, and the citations of social sciences in the year range of 2016–2020 (the third stage) are the highest. LISA and Geodetector are both related to the field of geography, and their citations of atmospheric science and meteorology are the most recent, which reflects that research about this discipline was developed later than other disciplines. Owing to the essential characteristics of the LISA model, it has been mostly applied in the social sciences, which is consistent with the LISA Geotree. LISA has been widely applied in health sciences since 2006 and has received more citations than geosciences studies, revealing that transdisciplinary research between geography and health has been growing rapidly.

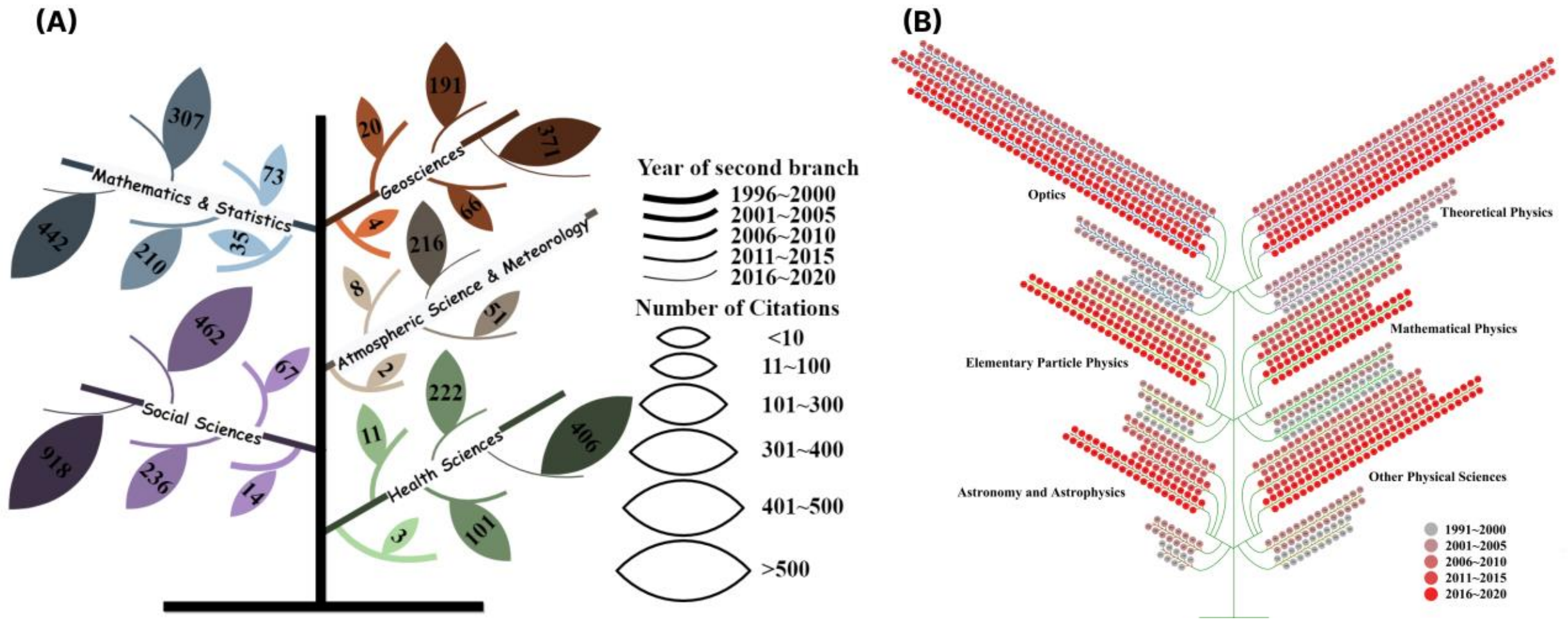

**Fig.7. Modeling of knowledge evolution trees for two cases in space science and physics**
**(A):** The Geotree of LISA. The branches represent five discipline types (Social Sciences: purple, Health Sciences: green gradient, Mathematics & Statistics: blue gradient, Atmospheric Science & Meteorology: grey gradient, and Geosciences: brown gradient), the twigs from root to top represent five different years ranging from old to new, distinguished by five levels of thicknesses, and six sizes of leaves represent different quantity levels of citations. The numbers on each leaf are the specific quantity of citations in that discipline in that year. **(B):** The Geotree of Special Relativity (SR). The first-level branches represent six classified disciplines; optics, theoretical physics, elementary particle physics, mathematical physics, astronomy and astrophysics, and other physical sciences; the second-level branches represent five time periods; and each leaf represents a citation of a paper on special relativity. Among them, the second-level branches farther from the root of the tree represent the citations in the newer year period.

### *3.3.3 The evolution tree of SR*

Similarly, the Geotree of Special Relativity (SR) was constructed (see **Fig.7 B**). On March 7, 2021, 2175 citations of the source publication were retrieved, which was titled "On the Electrodynamics of Moving Bodies," where Albert Einstein originally proposed the theory of SR on September 26, 1905 (Einstein, 1905). In the WoS database, a total of 1767 records for SR (1905–2021) were downloaded. Among them, 1281 records of SR included abstracts. After excluding citations in 2021, 1263 were used for the SR tree (from 1991 to 2020). All SR citations from 1991 to 2020 were divided into five stages. Each twig of each branch on the SR tree represents a stage of five and is marked in different colors. Special relativity involved six subject areas in terms of NSF codes: optics, theoretical physics, elementary particle physics, mathematical physics, astronomy and astrophysics, and other physical sciences, with 306, 312, 153, 182, 106, and 204 citations, respectively. Among them, research was mostly distributed in the fields of optics and theoretical physics in three stages: 2006–2010, 2011–2015, and 2016–2020. In contrast to other disciplines, the citations for Optics and Theoretical Physics in the fifth stage were less than those of the fourth stage, which reflected a "senescence stage." Furthermore, a period of "moth decay" was evident in mathematical physics, astronomy, and astrophysics from 2011 to 2015.

The knowledge development patterns of these three cases were distinct, and their application in different disciplines left gaps in quantity and nature. Geotree can be used to identify knowledge diffusion patterns and developmental gaps. Due to the scope of this study, an analysis of disciplinary gaps and diffusion mechanisms has not been carried out on the latter two theories. Based on the existing resulting trees, we found that the knowledge diffusion process was divided into stages of budding, growth, maturity, and senescence; some also had a stage of moth decay. In the budding period, the citation volume grew slowly from scratch; in the growing period, the citation growth rate was the fastest; and in the mature period, the citation growth rate tended to be stable. Taking LISA as an example, citation growth with a large time span may experience a "senescence stage," that is, the number of citations was lower than in the previous stage, and it appeared as a descending staircase on the evolution tree. Taking the special theory of relativity as an example, citation growth may have a period of “moth decay” in which it first increased, then decreased, and then increased, appearing as a sawtooth in the evolution tree.

## 4. Conclusions

Knowledge diffusion is a global phenomenon in society and academia (Kang, Kang, & Jang, 2023; Liang & Xu, 2023). Our literature review shows that the pathways and future trends of knowledge diffusion from the source to multiple disciplines, especially in specific fields, remained unclear due to limited studies. To better understand the process of knowledge diffusion, this study discussed the anatomy of knowledge evolution paths, tracing the transition of ideas from individuals to collective recognition. It explored how individual ideas evolved into shared knowledge, focusing on the critical elements and dynamics of this transformation. Utilizing the Geotree model and Geodetector case, our research started from citation growth to topic branches to branched evolution, future predictions, and their influencing factors. The findings revealed that from 2010 to 2020, the Geodetector theory’s spread underwent three distinct phases: an initial period of exploration, a subsequent growth period characterized by the emergence of numerous topics, and a mature phase marked by the fastest citation growth. The first paper contributed most significantly to citation growth, while subsequent publications contributed to the transition from the embryonic to the mature stage of citations and knowledge. Our research also uncovered two additional stages from other cases, the moth-eaten stage, and the aging stage, which are potential future trends in the advancement of Geodetector knowledge. By 2030, this growth trend is projected to drive more practical applied articles, while theoretical developmental articles are expected to remain at a low level. Geodetector’s knowledge has disseminated from Asia in 2010 to five other continents in different years, resulting in globally distributed multidisciplinary applications. This global distribution of Geodetector knowledge was significantly explained by the overall publishing capabilities of economies (q=0.307***). In turn, this domain-specific knowledge diffusion affected the knowledge distribution across all fields, particularly during its growth stage (q-statistic = 0.151***), highlighting its global influence and multidisciplinary applications.

For those scholars who are using, or want to apply Geotree in knowledge diffusion, it should be noted that in previous evolution tree-related studies (Duan et al., 2020; Jing et al., 2022; J. F. Wang et al., 2012): the primary branches of the evolutionary tree denoted type variables; secondary branches denoted time variables; leaves represented research unit; the tree structure reflected the characteristics of all research units in different classifications and stages of development; and the total number of research units under each stage remained unchanged, such as the total number of countries in the world. However, since our input was citation data under different research topics, the type and stage divisions of citation documents varied with the research case, and the total number of citations increased over time. Therefore, the evolutionary tree structure modeling in this paper was developed in a similar way but in a different context, with reference to previous Geotree cases and knowledge diffusion characteristics. Each leaf in the three tree models represented a different research object, namely a citation, a research direction, and a country. Among them, the first one had both primary and secondary branches indicating the topic classification and dynamic pathways, so we draw the discipline-research direction tree for each year from 2010 to 2020; the second one reflected the evolution characteristics of citations under different stages and different topics, and it should be noted that the total number of leaves under each stage changed with the growth of citations; the third case introduced country types and development stages in the growth of citations, in the same mode as traditional geotrees.

In conclusion, our research significantly advanced the field by filling critical gaps in the existing literature, offering a robust framework for analyzing knowledge diffusion, and providing actionable insights for future research and policy-making. The innovative use of Geotree modeling demonstrated the potential of integrated tools in addressing complex research questions, setting a new standard for studies in knowledge diffusion.

## 5. Implications and limitations

These results have both theoretical and practical implications.

Theoretically, unlike previous scholars who have adopted methods like bibliometrics, econometrics, and mathematical statistics (Barnett, Huh, Kim, & Park, 2011; Haeussler & Sauermann, 2020; Kiss et al., 2010), which revealed scientific advances and knowledge diffusion of the specified topics, keywords, and journals in search terms. This study innovatively traced the process of knowledge diffusion from representative publications of a scientific idea, introducing a methodology that enabled modeling, prediction, and analysis of influencing factors, expressing the multiple dimensions of knowledge diffusion: spatial, temporal, and thematic. Specifically, distinguishing from the previous visualizations and structures (Ba et al., 2023; X. Liu, Zhao, Wei, & Abedin, 2024), the tree-like hierarchy provided a new way in this field from the perspective of ecology to express the knowledge growth dynamics. So, we integrated the branching information from the tree modeling to improve the predicting performance. Compared with the previous studies using proximal variables from the citation database (e.g., the length of the Abstract, the linguistic fearures) (K. Chen, Song, Zhao, Peng, & Chen, 2024; Didegah & Thelwall, 2013; Wagner et al., 2021), we introduced social economic and scientific productivity factors to analyze the global distribution of knowledge dissemination. In short, this holistic approach contrasted with existing studies, enabling a unified analysis from data collection to methodological development and explanatory insights.

In practice, by analyzing the multidimensional spread of knowledge from its initial source to the practical applications, the potential practical implications are as follows: Firstly, niche domains have gradually generated citing linkages across research directions since their inception of knowledge (Haeussler & Sauermann, 2020; A. M. Petersen, D. Majeti, K. Kwon, M. E. Ahmed, & I. Pavlidis, 2018). Such linkages beyond boundaries expanded the breadth of scientific influence in small fields, aided in the breakthrough in the communication of interdisciplinary theories, and created new research avenues (Bu et al., 2021; A. M. Petersen et al., 2018; Trujillo & Long, 2018). Secondly, knowledge diffusion evolves from emergence to growth and maturity, possibly entering the moth-eaten and aging stages. At a stage of knowledge depletion and stagnant growth, strategies such as leveraging complementary assets (e.g., releasing software and publishing papers) can spur new growth by enhancing diffusion rates and topic diversity (Didegah & Thelwall, 2013; Methé, 1992; Wagner et al., 2021). Our findings also show that GDP is not a primary factor in adopting knowledge; however, the scientific and engineering (S&E) article output of economies accounts for 30.7% of the variance in knowledge adoption rates. In turn, the proliferation of knowledge of the Geodetector theory at each stage contributed to approximately 15% of the S&E article output. This suggested that each economy's capacity to produce S&E articles in general influenced scholars' application in specific fields, and its long-term consequences of knowledge diffusion and integration in turn played an important role in multidisciplinary collaborations.

However, despite the contributions made by this study, this research is not without its limitations. Future research can enhance our findings in several ways: (1) Considering the academic quality of publications, we collected data from the WoS database, but it should be noted that Geodetector also covered some citing records that were not included in WoS. More complete citation data can be integrated through different databases such as Google Scholar and Scopus. (2) For the thematic analysis of the citation data, we applied the commonly-used LDA algorithm to define each document with one unique discipline type and research direction type. To improve classification accuracy, an integrating approach combining AI and WoS categories could be explored in the future. (3) Aside from GDP, research and development (R&D) is also closely related to knowledge diffusion and can indicate the strength of investment in research (Savin and Egbetokun, 2016). The growth of R&D and S&E is closely related, with global R&D spending surging from $722 billion in 2000 to $2.14 trillion in 2017. In tandem, the global production of S&E papers has increased from 990,000 in 1996 to 2.94 million in 2020. To reduce the bias of research data and construct a more representative knowledge factor evolution tree, the impact of R&D inputs should be incorporated through treatments such as standardization and removal of covariance (H. Kim & Park, 2009; Sanso-Navarro & Vera-Cabello, 2018). Through the existing research framework, we have quantified the branching paths of knowledge diffusion over time and its future trends. The feasibility of this approach was also demonstrated by two other cases. Therefore, conducting targeted case studies to examine the heterogeneous diffusion of tools, theories, and ideas within academia is feasible. These dissemination paths branched out depending on time, space, and research interests. Such investigations not only contribute to the development of scientific advancements but also foster interdisciplinary collaboration and the potential creation of new knowledge.

**Acknowledgments**

We acknowledge support from the EU-funded TREND project (Horizon 2020 Excellent Science - Marie Skłodowska-Curie Actions), Grant ID: 823952.

**Funding:** This study was funded by the National Natural Science Foundation of China (Grant No. 42071375), the Innovation Project of State Key Laboratory of Resources and Environmental Information System (Grant No. O88RA205YA), and the China Scholarship Council (Grant No. 202204910278).

**Competing interests:** Authors declare that they have no competing interests.

**Data and materials availability:** All data are available in the main text.

**Ethical approval:** This article does not contain any studies with human participants performed by any of the authors

**Informed consent:** This article does not contain any studies with human participants performed by any of the authors